\RequirePackage{fixltx2e}
\documentclass[twocolumn,floatfix,showpacs,preprintnumbers,nofootinbib,%
superscriptaddress]{revtex4}
\usepackage[english]{babel}
\usepackage[utf8x]{inputenc}
\usepackage{mathrsfs}
\usepackage{amssymb}
\usepackage{amsmath}
\usepackage{mathtools}
\usepackage{mathrsfs}
\usepackage{graphicx}
\usepackage[caption=false]{subfig}
\usepackage{fixltx2e}

\usepackage{xcolor}
\definecolor{lcolor}{rgb}{0.5,0,0}
\definecolor{citcolor}{rgb}{0,0.3,0.0}

\usepackage[breaklinks,colorlinks,urlcolor=blue,citecolor=citcolor,linkcolor=lcolor]{hyperref}
\usepackage{subdepth}

\newcommand{\der}{\mathrm{d}}

\newcommand{\tr}{\, \mathrm{Tr} \, }

\newcommand{\nc}{{N_\mathrm{c}}}
\newcommand{\nf}{{n_\mathrm{f}}}

\newcommand{\gev}{\ \textrm{GeV}}

\newcommand{\qs}{Q_\mathrm{s}}

\newcommand{\qso}{Q_\mathrm{s,0}}

\newcommand{\lqcd}{\Lambda_{\mathrm{QCD}}}
\newcommand{\as}{\alpha_{\mathrm{s}}}
\newcommand{\asbar}{\bar{\alpha}_{\mathrm{s}}}

\newcommand{\fig}{Fig.~}

\newcommand{\eq}{Eq.~}
\newcommand{\se}{Sec.~}

\newcommand{\re}{Ref.~}
\newcommand{\res}{Refs.~}

\begin{document}

\author{T. Lappi}
\affiliation{
Department of Physics, University of Jyv\"askyl\"a %
 P.O. Box 35, 40014 University of Jyv\"askyl\"a, Finland
}

\affiliation{
Helsinki Institute of Physics, P.O. Box 64, 00014 University of Helsinki,
Finland
}
\author{H. M\"antysaari}
\affiliation{
Physics Department, Brookhaven National Laboratory, Upton, NY 11973, USA
}

\title{
Next-to-leading order Balitsky-Kovchegov equation with resummation
}

\pacs{
12.38.Cy      
}

\preprint{}

\begin{abstract}
We solve the Balitsky-Kovchegov evolution equation at next-to-leading order accuracy including a resummation of large single and double transverse momentum logarithms to all orders. We numerically determine an optimal value for the constant under the large transverse momentum logarithm that enables including a maximal amount of the full NLO result in the resummation. When this value is used the contribution from the $\as^2$ terms without large logarithms is found to be small at large saturation scales and at small dipoles. Close to  initial conditions relevant for phenomenological applications these fixed order corrections are shown to be numerically important.
\end{abstract}

\maketitle

\section{Introduction}

In high energy hadronic collisions perturbative QCD predicts a rapid growth of  gluon densities, as emissions of gluons that carry a small longitudinal momentum fraction are favored. At such high densities  non-linear saturation phenomena become important. The Color Glass Condensate (CGC)~\cite{Gelis:2010nm} has proven itself to be a powerful effective field theory to describe the strong interactions in these high-density environments. Leading order CGC calculations have been able to successfully describe qualitatively, and also semi-quantitatively, many high-energy scattering processes where the small-$x$ (longitudinal momentum fraction) part of the hadronic wave function is probed. These include, for example, deep inelastic scattering~\cite{Albacete:2010sy} and single~\cite{Albacete:2010bs,Tribedy:2011aa,Rezaeian:2012ye,Lappi:2013zma} and double inclusive~\cite{Albacete:2010pg,Stasto:2011ru,Lappi:2012nh,
JalilianMarian:2012bd} particle production. The CGC framework has also been successfully applied to calculations of the initial state for hydrodynamical modeling of a heavy ion 
collision~\cite{Lappi:2011ju,Schenke:2012wb,Gale:2012rq}.

When describing high-energy scattering in QCD it is useful to employ the eikonal approximation. The most convenient degrees of freedom are then the transverse coordinate dependent Wilson lines that describe the eikonal propagation of a quark or a gluon trough the dense color field of the target. Cross sections can be expressed in terms of correlators of Wilson lines, the most simple one being the dipole (correlator of two fundamental representation Wilson lines) which gives the scattering amplitude for the quark-antiquark dipole to scatter off a hadronic target. A necessary ingredient in many CGC calculations of cross sections is  the Balitsky-Kovchegov (BK) equation, which determines the dependence of this dipole amplitude on rapidity (or, equivalently, in Bjorken-$x$ or energy). It was first derived at leading order in \res\cite{Balitsky:1995ub,Kovchegov:1999yj} and at next-to-leading order in \re\cite{Balitsky:2008zza}.

When perturbative QCD calculations are done in the collinear factorization framework, next-to-leading order (NLO) corrections are known to be numerically significant. The same could be expected also in the CGC. Thus, in order to test our understanding of  saturation phenomena encountered in  high-energy collisions, the CGC calculations must be made more quantitative by calculating  the cross sections at NLO accuracy. First steps in this direction have been taken recently by calculating the single inclusive~\cite{Altinoluk:2011qy,Chirilli:2011km,Chirilli:2012jd,Stasto:2013cha,Altinoluk:2014eka} and DIS cross sections~\cite{Balitsky:2010ze,Beuf:2011xd} at this order in the QCD coupling $\as$. However, it is not consistent to use these the NLO cross section calculations without a solution to the corresponding  NLO evolution equation. 

The NLO BK equation was solved numerically for the first time recently in \re\cite{Lappi:2015fma}. Its linearized version, the NLO~BFKL equation has been known before~\cite{Fadin:1996nw,Fadin:1998py,Ciafaloni:1998gs}, and a solution to it with an absorptive boundary conditions (to mimic the non-linear effects) also exists~\cite{Avsar:2011ds}. 
The NLO~BFKL equation includes large logarithms of transverse momentum that have been resummed in \res\cite{Salam:1998tj,Ciafaloni:1999yw,Altarelli:1999vw,Ciafaloni:2003rd}. However, as the BFKL equation is valid only in the linear regime where the scattering amplitude is small, these resummations can not be straightforwardly applied to the BK equation.  For the non-linear BK equation a resummation scheme for the large transverse logarithms has been developed recently~\cite{Iancu:2015vea,Iancu:2015joa}. In addition to these resummations, there have also been proposals to include a kinematical constraint in the BK equation~\cite{Motyka:2009gi,Beuf:2014uia}.

In our previous publication~\cite{Lappi:2015fma} we showed that the NLO BK equation does not always give a physically meaningful evolution, and can not be applied to phenomenology. In this work, we study how the resummation of large transverse logarithms proposed in \res\cite{Iancu:2015vea,Iancu:2015joa} changes this picture.

This paper is organized as follows. First, in \se\ref{sec:nlobk} we briefly review the NLO BK equation, and present the resummation of large logarithms to the equation in \se\ref{sec:resum}.
The numerical solution of the resummed NLO evolution equation is discussed in \se\ref{sec:results}.

\section{Balitsky-Kovchegov equation at next to leading order}
\label{sec:nlobk}
The Balitsky-Kovchegov equation describes the rapidity evolution of the dipole operator which can be written as a correlator of two Wilson lines $U$:
\begin{equation}
	S(x-y) = \frac{1}{\nc} \langle \tr(U_x U^\dagger_y) \rangle .
\end{equation}
Here the brackets $\langle \rangle$ refer to an average over the target color field and $x$ and $y$ are transverse coordinates. The dependence on rapidity (or Bjorken-$x$) of the Wilson lines is left implicit. The next-to-leading order evolution equation for the dipole operator in rapidity can be written as:
\begin{multline}
\label{eq:nlobk}
\partial_y \frac{1}{\nc}  \tr (U_x U^\dagger_y)  = \frac{\as \nc}{2\pi^2} K_1^\text{BC} \otimes D_1 \\
+ \frac{\as^2 \nc^2}{8\pi^4} K_2 \otimes D_2 + \frac{\as^2 \nf\nc}{8\pi^4} K_f \otimes D_f.
\end{multline}
The kernels and Wilson line operators derived in~\cite{Balitsky:2008zza} are
\begin{widetext}
\begin{align}
\label{eq:orig_k1}
K^\textnormal{BC}_1 &= \frac{r^2}{X^2Y^2} \left[ 1+\frac{\as\nc }{4\pi} \left(  \frac{\beta}{\nc} \ln r^2\mu^2 - \frac{\beta}{\nc} \frac{X^2-Y^2}{r^2} \ln \frac{X^2}{Y^2} + \frac{67}{9} - \frac{\pi^2}{3} - \frac{10}{9} \frac{\nf}{\nc} - 2\ln \frac{X^2}{r^2} \ln \frac{Y^2}{r^2} \right) \right]
\\
D_1 &= \frac{1}{\nc} \tr (U_x U^\dagger_z) \frac{1}{\nc} \tr(U_z U^\dagger_y) - \frac{1}{\nc} \tr (U_x U^\dagger_y)  \\
K_2 &= -\frac{2}{(z-z')^4} + \left[ \frac{X^2 Y'^2 + X'^2Y^2 - 4r^2(z-z')^2}{(z-z')^4(X^2Y'^2 - X'^2Y^2)} + \frac{r^4}{X^2Y'^2(X^2Y'^2 - X'^2Y^2)} + \frac{r^2}{X^2Y'^2(z-z')^2} \right]\\
& \quad  \times  \ln \frac{X^2Y'^2}{X'^2Y^2} \nonumber
\\
D_2 &= \frac{1}{\nc} \tr (U_x U^\dagger_z) \frac{1}{\nc}  \tr (U_z U^\dagger_{z\smash{'}}) \frac{1}{\nc} \tr (U_{z\smash{'}}U^\dagger_y)  - \frac{1}{\nc} \tr (U_x U^\dagger_z ) \frac{1}{\nc}  \tr (U_z U^\dagger_y) 
\\ \nonumber
& \quad - \frac{1}{\nc^3} \tr (U_x U^\dagger_z U_{z\smash{'}} U^\dagger_y U_z U^\dagger_{z\smash{'}}) + \frac{1}{\nc^3} \tr (U_x U^\dagger_y) \\
 K_f &= \frac{2}{(z-z')^4}  
	- \frac{X'^2Y^2 + Y'^2 X^2 - r^2 (z-z')^2}{(z-z')^4(X^2Y'^2 - X'^2Y^2)} \ln \frac{X^2Y'^2}{X'^2Y^2} \\
D_f &= \frac{1}{\nc} \tr (U_y U^\dagger_z) \left( \frac{1}{\nc} \tr (U_x U^\dagger_{z\smash{'}}) - \frac{1}{\nc} \tr (U_x U^\dagger_z)\right) + \frac{1}{\nc^4} \tr (U_xU^\dagger_y)\tr( U_z U^\dagger_{z\smash{'}}) 
\\ \nonumber
	&\quad - \frac{1}{\nc^3} \tr( U_x U^\dagger_y U_z U^\dagger_{z\smash{'}}) - \frac{1}{\nc^3} \tr (U_x U^\dagger_{z\smash{'}} U_z U^\dagger_y)  + \frac{1}{\nc^3} \tr (U_xU^\dagger_y)
\end{align}
\end{widetext}
The convolutions $\otimes$ in \eq\eqref{eq:nlobk} denote integration over the transverse coordinate $z$ (in $K_1^\text{BC}$) or $z$ and $z'$ (in $K_2$ and $K_f$). We use the notation $X^2=(x-z)^2$, $X'^2=(x-z')^2$, $Y=(y-z)^2$ and $Y'=(y-z')^2$. 

Because every trace is proportional to $\nc$, in the large-$\nc$ limit the terms with traces of more than two Wilson lines can be neglected. The large-$\nc$ limit also implies the mean-field limit, where the correlators of products of traces factorize into products of the two-point function $S(r)$. This mean-field limit closes the equation: the rapidity derivative of the dipole operator $S(r)$ can be computed in terms of $S(r)$ only. 
At finite $\nc$, correlators of more than two Wilson lines are needed which, in principle, have their own evolution equations. In that case one should solve an infinite hierarchy of coupled evolution equations, or equivalently the JIMWLK~\cite{JalilianMarian:1996xn,JalilianMarian:1997jx, JalilianMarian:1997gr,Iancu:2001md, Ferreiro:2001qy, Iancu:2001ad, Iancu:2000hn} equation at NLO accuracy~\cite{Balitsky:2013fea,Kovner:2013ona}. This would be numerically demanding, and a much more practical approach could be to use e.g. the so called Gaussian approximation (see e.g. \re\cite{Dominguez:2011wm}) to express the higher-point functions in terms of the dipole operator only. As the effect of the finite-$\nc$ corrections to the leading order BK equation is known to be much smaller than $\sim 1/\nc^2\approx 10\%$ (which would be a naive expectation from the $1/\nc$ expansion)~\cite{Kovchegov:2008mk}, we take the large-$\nc$ limit in this work. 

One of the NLO corrections is the running of the QCD coupling $\as$. The term involving the renormalization scale $\mu^2$ in \eq\eqref{eq:orig_k1} should be absorbed into the running of $\as$. What other terms are included in the scale-dependent coupling is a scheme choice.  We adopt the choice derived in Ref.~\cite{Balitsky:2006wa} and replace all terms in $K_1^\text{BC}$ proportional to the beta function coefficient $\beta=\frac{11}{3}\nc - \frac{2}{3}\nf$ (with $\nf=3$ in this work) by the so called Balitsky running coupling.
This  prescription is used here because we want to resum all large logarithms, and the Balitsky running coupling resums $\as \beta$ contributions, and in particular the logarithm $\sim \beta \ln X^2/Y^2$ from $K_1^\text{BC}$.

The Balitsky prescription has been successfully used in phenomenological applications to include running coupling effects in the leading order BK equation. For the other terms we choose to evaluate $\as$ at the scale given by the size of the parent dipole $r$, as it is the only available external scale. Notice also that for the $\as^2$ terms the difference between the scale choices for the coupling is formally a higher-order $\as^3$ correction. The kernel $K_1$ can now be written as
\begin{widetext}
\begin{multline}
	\frac{\as \nc}{2\pi^2} K_1^\textnormal{Bal} = \frac{\as(r) \nc}{2\pi^2} \left[\frac{r^2}{X^2Y^2} + \frac{1}{X^2} \left(\frac{\as(X)}{\as(Y)}-1\right) + \frac{1}{Y^2} \left(\frac{\as(Y)}{\as(X)}-1\right) \right] \\
		+ \frac{\as(r)^2 \nc^2}{8\pi^3} \frac{r^2}{X^2Y^2} \left[ \frac{67}{9} - \frac{\pi^2}{3} - \frac{10}{9} \frac{\nf}{\nc} - 2\ln \frac{X^2}{r^2} \ln \frac{Y^2}{r^2} \right].
\end{multline}
\end{widetext}

The strong coupling constant $\as$ at the given distance scale $r$ is evaluated as
\begin{equation}
  \as(r) = \frac{4\pi}{
\beta \ln\left\{\left[
       \left(\frac{\mu_0^2}{\lqcd^2}\right)^{\frac{1}{c}}
      +\left(\frac{4e^{-2 \gamma_\mathrm{E}}}{r^2\lqcd^2}\right)^{\frac{1}{c}} \right]^{c}
\right\} }.
\end{equation}
The parameters $c$ and $\mu_0$ control the infrared behavior of the coupling constant, and here we take $\mu_0/\lqcd=2.5$ and $c=0.2$, which freezes the coupling to $\approx 0.76$ in the infrared.
Note the constant factor $4e^{-2\gamma_\mathrm{E}}\approx 1.26$ in the identification 
$k^2 \sim 4e^{-2\gamma_{\mathrm{E}}} /r^2$, which is taken from the explicit
Fourier transform of the kernel calculated analytically in \res\cite{Gardi:2006rp,Kovchegov:2006vj} and confirmed numerically in \re\cite{Lappi:2012vw}. In the leading order fits to the deep inelastic scattering data the scale at which the coupling is evaluated is taken as a fit parameter by identifying $k^2 \sim 4C^2/r^2$. These fits require $C^2\sim 4 \dots 20$ in order to get a slow enough evolution speed~\cite{Albacete:2010sy}. 
In this work we do not seek parametrizations that give a best fit to the DIS data, and use the theoretically motivated value $C^2=e^{-2\gamma_\mathrm{E}}$.

The NLO BK equation was first solved in \re\cite{Lappi:2015fma} where it was shown that the equation is unstable. In particular, depending on the initial condition the dipole amplitude $N(r)=1-S(r)$ may decrease or even become negative when rapidity increases, which is unphysical as it would correspond to a decrease of the  unintegrated gluon distribution when decreasing the momentum fraction $x$. The origin of this problematic behavior was traced back to the double logarithmic term $\sim \ln \frac{X^2}{r^2} \ln \frac{Y^2}{r^2}$ in the kernel $K_1$. To fix this problem, a resummation of large logarithmic corrections is needed.

\section{Resumming large logarithms}
\label{sec:resum}
There are two sources of large logarithmic corrections to the BK equation that must be resummed to all orders. First, as shown in \re\cite{Iancu:2015vea}, the successive gluon emissions that are strongly ordered in both transverse and longitudinal momenta
generate a large double logarithmic contribution $\sim \ln X^2/r^2 \ln Y^2/r^2$ to the NLO BK equation. 
These contributions are resummed in \re\cite{Iancu:2015vea} to all orders in $\as \ln X^2/r^2 \ln Y^2/r^2$, and the effect of the resummation is to remove the double logarithmic term from the kernel $K_1$, and multiply it by an oscillatory factor
\begin{equation}
	K_\text{DLA}=\frac{J_1\left(2\sqrt{\asbar x^2}\right)}{\sqrt{\asbar x^2}} \approx 1 - \frac{ \asbar x^2}{2} + \mathcal{O}(\asbar^2).
\end{equation}
The double logarithm here is $x=\sqrt{\ln X^2/r^2 \ln Y^2/r^2}$, and $\asbar = \as \nc/\pi$. If $\ln X^2/r^2 \ln Y^2/r^2<0$, then an absolute value is used and the Bessel function is changed to $J_1 \to I_1$, see \re\cite{Iancu:2015vea}.

In addition to the kernel of the evolution equation, also the initial condition for the BK evolution must be resummed. For this the dipole amplitude is parametrized as
\begin{equation}
\label{eq:ic}
	N(r) = 1 - \exp \left(-r^2 \qso^2 \tilde A(\rho)\right),
\end{equation}
where $\rho = \ln 1/(r^2\qso^2)$ and the parameter $\qso$ controls the value of the saturation scale $\qs$ at the initial condition. The resummed factor $\tilde A$ is obtained from the original $A$ as
\begin{equation}
	\tilde A(\rho) = \int_0^\rho \der \rho_1 \Big[\delta(\rho-\rho_1) - \sqrt{ \asbar} J_1(2\sqrt{\bar \as  (\rho-\rho_1)^2}\Big]A(\rho_1).
\end{equation}
The McLerran-Venugopalan (MV) model~\cite{McLerran:1994ni}  corresponds to $A(\rho)=\rho$, which gives
\begin{equation}
\begin{split}
\tilde A(\rho) = \frac{\rho}{2} \Big[ 1 + J_0(2\sqrt{\as \rho^2}) + \frac{\pi}{2}H_0(2\sqrt{\as \rho^2})J_1(2\sqrt{\as \rho^2}) \\
- \frac{\pi}{2} H_1(2\sqrt{\as \rho^2})J_0(2\sqrt{\as \rho^2}) \Big].
\end{split}
\end{equation}
In order to obtain a dipole amplitude that has a correct behavior in the infrared limit we include also an infrared cutoff and replace the prefactor $\rho/2$ by $\ln (1/r\qso + e)$. Note that this parametrization is not exactly the MV model used in  our previous work~\cite{Lappi:2013zma}, but we choose to use it here in order to be consistent with \re\cite{Iancu:2015vea}.

There is also a large single transverse logarithm (STL) in the evolution equation that forbids us to do only a simple $\as$ expansion. As shown in \re\cite{Iancu:2015joa}, the large transverse logarithm $\sim \ln (1/r\qs)$ at the order $\as^2$ originates from the kernel $K_2$, namely from the part
\begin{multline}
\label{eq:mstl}
M_\text{STL} = -\frac{2}{(z-z')^4} \\
+ \frac{X^2Y'^2 + X'^2Y^2 - 4r^2(z-z')^2}{(z-z')^4(X^2Y'^2-X'^2Y^2)} \ln \frac{X^2Y'^2}{X'^2Y^2}.
\end{multline}
Note that the other terms in the kernels $K_2$ and $K_f$ are suppressed by powers of $r^2$ in the small dipole limit.
These large logarithms (at small $r$) appear together with $\as$ at all orders and can also be resummed. The resummation was done in  \re\cite{Iancu:2015joa} by multiplying the kernel $K_1$ by a factor
\begin{equation}
	\label{eq:k_stl}
	K_\text{STL} = \exp \left\{ - \frac{\as \nc A_1}{\pi} \left| \ln \frac{C_\text{sub} r^2}{\min\{X^2,Y^2\}} \right| \right\}.
\end{equation} 
The leading logarithm resummation done in \re\cite{Iancu:2015joa} does not fix the constant factor $C_\text{sub}$ (which should be of the order one) in $K_\text{STL}$. We shall fix this coefficient later in such a way that the resummation captures as accurately as possible the full small-$r$ limit of $M_\text{STL}$, i.e. not only the leading logarithm. The constant $A_1=11/12$  comes from the DGLAP anomalous dimension for $q \to qg$ and $g \to gg$ splittings. Because the $\as^2$ part of this resummation is included in the kernel $K_2$, in order to avoid double counting we subtract the $\as^2$ piece of the single logarithm resummation $K_\text{STL}$ from the modified kernel $K_1$. 

With all these building blocks, we can write the kernel $K_1$ used in this work. It is obtained from the kernel of the NLO BK equation by including the Balitsky running coupling and resumming the large single and double transverse logarithms. Thus the final kernel used in the numerical calculation now reads
\begin{multline}
\label{eq:full_k1}
	\frac{\as \nc}{2\pi^2} K_1 = \frac{\as(r) \nc}{2\pi^2} K_\text{DLA} K_\text{STL} \\
\times
	 \left[\frac{r^2}{X^2Y^2} + \frac{1}{X^2} \left(\frac{\as(X)}{\as(Y)}-1\right) + \frac{1}{Y^2} \left(\frac{\as(Y)}{\as(X)}-1\right) \right] \\
	 - K_\text{sub} + K_1^\text{fin}.
\end{multline}
Here $K_\text{sub}$ subtracts the $\as^2$ part of the single transverse logarithm $K_\text{STL}$ which is included exactly in $K_2$. This subtraction term reads
\begin{equation}
	K_\text{sub}=\frac{\as(r) \nc}{2\pi^2} \left(-\frac{\as(r) \nc A_1}{\pi} \left| \ln \frac{C_\text{sub} r^2}{\min\{X^2,Y^2\}} \right| \right)  \frac{r^2}{X^2Y^2}.
\end{equation}
 Note that we choose to use the parent dipole running coupling in $K_\text{sub}$ as we want it to cancel the corresponding contribution originating from the kernel $K_2$ (the $M_\text{STL}$ part) which uses the same parent dipole prescription. 
Thus the subtraction term $K_\text{sub}$ does not precisely cancel the $\as^2$ term of the expansion of $K_\text{STL}$ times the full Balitsky running coupling. Because the difference between the running coupling schemes is of higher order in $\as$, $K_\text{sub}$ does, however, cancel the contribution from the $\as^2$ term in $K_\text{STL}$ to the order $\as^2$, which is enough for the purpose of this work.
The other NLO terms in $K_1$ that are not included in the resummation are denoted by $K_1^\text{fin}$, which reads
\begin{equation}
K_1^\text{fin} = \frac{\as(r)^2 \nc^2}{8\pi^3} \frac{r^2}{X^2Y^2} \left[ \frac{67}{9} - \frac{\pi^2}{3} - \frac{10}{9} \frac{\nf}{\nc}  \right].
\end{equation}

\begin{figure}[tb]
\begin{center}
\includegraphics[width=0.49\textwidth]{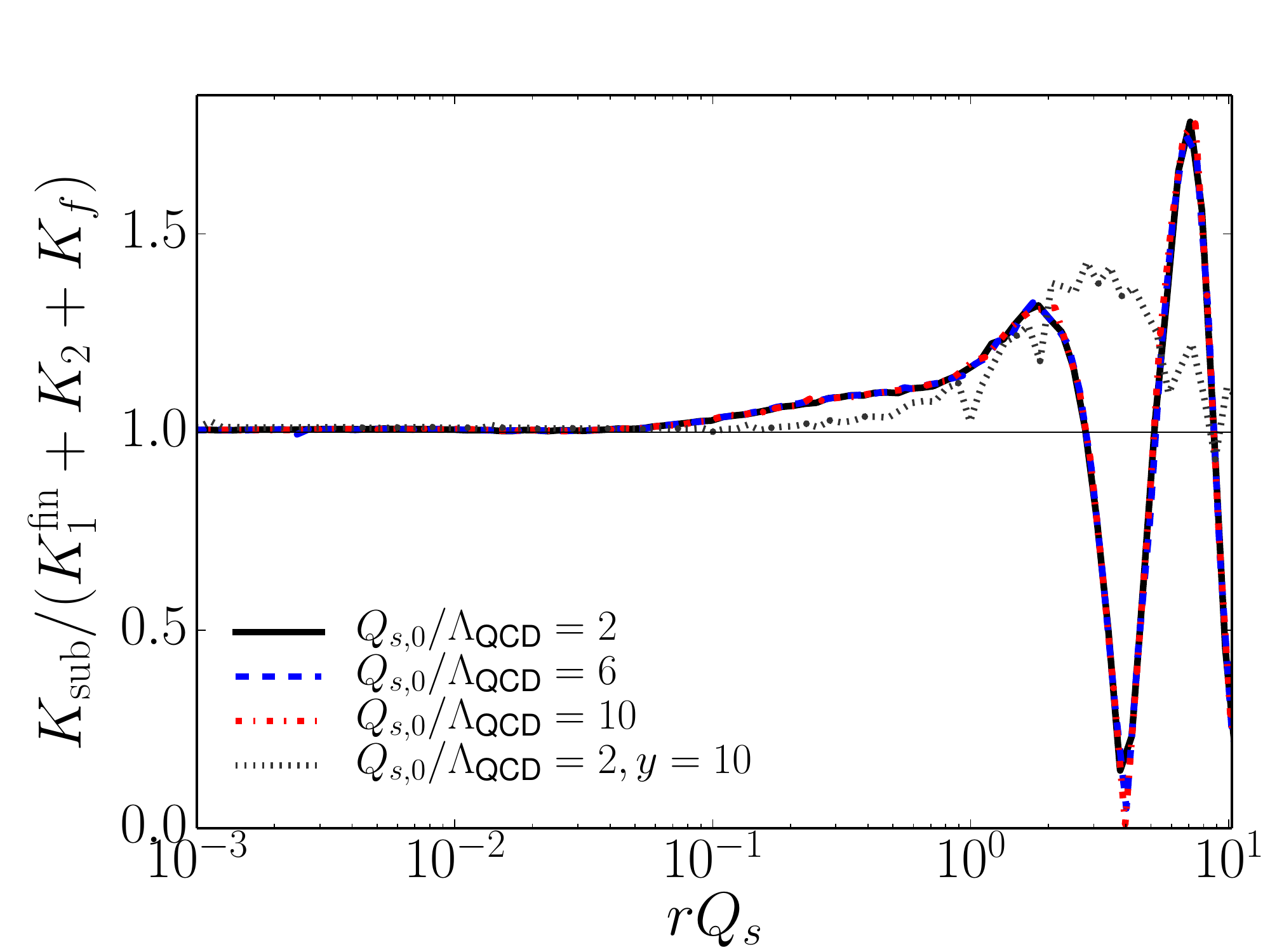}
\end{center}
\caption{
Contribution to the evolution speed of the dipole amplitude, $\partial_y N$, originating from the subtraction of the $\as^2$ part of the single logarithm resummation ($K_\text{sub}$) divided by the contribution from $K_1^\text{fin}$, $K_2$ and $K_f$. 
}\label{fig:dndy_resummation_subtraction}
\end{figure}

The subtracted $\as^2$ contribution $K_\text{sub}$ is a leading logarithm result, and it corresponds to the leading logarithmic behavior of the contribution from $K_2$. In order to include most of the next-to-leading order corrections into the resummation $K_\text{STL}$, we fix the constant $C_\text{sub}$ in \eq\eqref{eq:k_stl} by requiring that the subtraction term $K_\text{sub}$ reproduces as accurately as possible the small-$r$ limit of the other NLO terms. This procedure for determining $C_\text{sub}$ is demonstrated in 
\fig\ref{fig:dndy_resummation_subtraction}, where we plot the contribution to the rapidity derivative of the dipole amplitude $\partial_y N(r)$ from the subtraction term $K_\text{sub}$  divided by the contribution from kernels $K_1^\text{fin}$, $K_2$ and $K_f$ (convoluted with the corresponding dipole operators). The ratio is found to be close to unity within a wide range of parent dipole  sizes at the initial condition by choosing $C_\text{sub}=0.65$. For another value of $C_\text{sub}$ the ratio would approach unity only very slowly when the leading logarithm of $r$ numerically dominates the other terms. Thus, with $C_\text{sub}=0.65$, $K_\text{STL}$ includes as accurately as possible the leading small-$r$ part of the kernels $K_1^\text{fin}$, $K_2$ and $K_f$. We regard this choice as ``optimal'' in the sense that it includes a maximal part of the small-$r$ NLO contribution in the (numerically easier) resummation. Thus such a choice minimizes the contribution of the numerically more difficult non-logarithmic other NLO contributions (that were neglected in 
Ref.~\cite{Iancu:2015joa}). The ratio is also shown for $\qso/\lqcd=2$ after $10$ units of rapidity evolution, and it can be seen that the subtraction term is still capturing most of the NLO corrections with the same $C_\text{sub}$. We have checked that modifying the value of $C_\text{sub}$ within a factor of $2$ moves contributions between the resummation and $\as^2$ terms without significantly affecting the overall evolution.

\begin{figure}[tb]
\begin{center}
\includegraphics[width=0.49\textwidth]{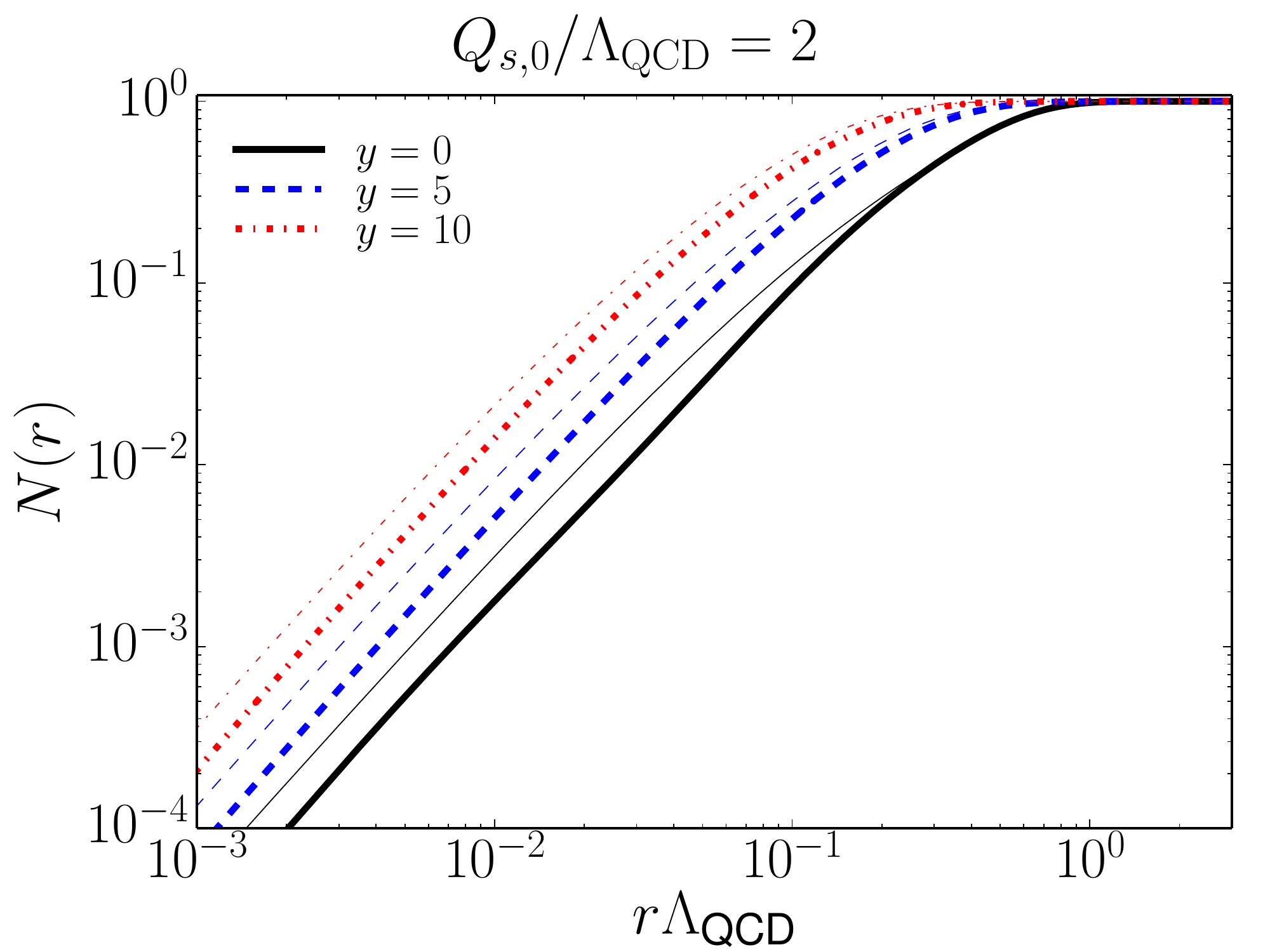}
\end{center}
\caption{
Dipole amplitude at different rapidities as a function of dipole size. The thick lines are obtained by using a resummed initial comparison. For comparison, the corresponding amplitudes obtained without resumming the initial condition are shown as thin lines.
}\label{fig:amplitude}
\end{figure}

\section{Evolution of the dipole amplitude}
\label{sec:results}
The dipole amplitudes $N(r)=1-S(r)$ at rapidities  $y=0,5$ and $y=10$ obtained by solving the resummed NLO BK equation are shown in \fig\ref{fig:amplitude}. 
The amplitude is found to increase at almost all dipole sizes through the evolution. In particular, the amplitude does not turn negative at small dipoles, which would be the case with the NLO BK equation without resummation as shown in \re\cite{Lappi:2015fma}. In order to study the effect of the resummed initial condition we also solve the equation with a non-resummed dipole amplitude at $y=0$ (replacing $\tilde A$ by $A$ in \eq\eqref{eq:ic}).
The difference between the initial conditions is that the resummation introduces oscillations in the small-$r$ part that are quickly washed out in the evolution. The evolution speeds and shapes of the solutions are comparable after a few units of rapidity evolution. 

\begin{figure}[tb]
\begin{center}
\includegraphics[width=0.49\textwidth]{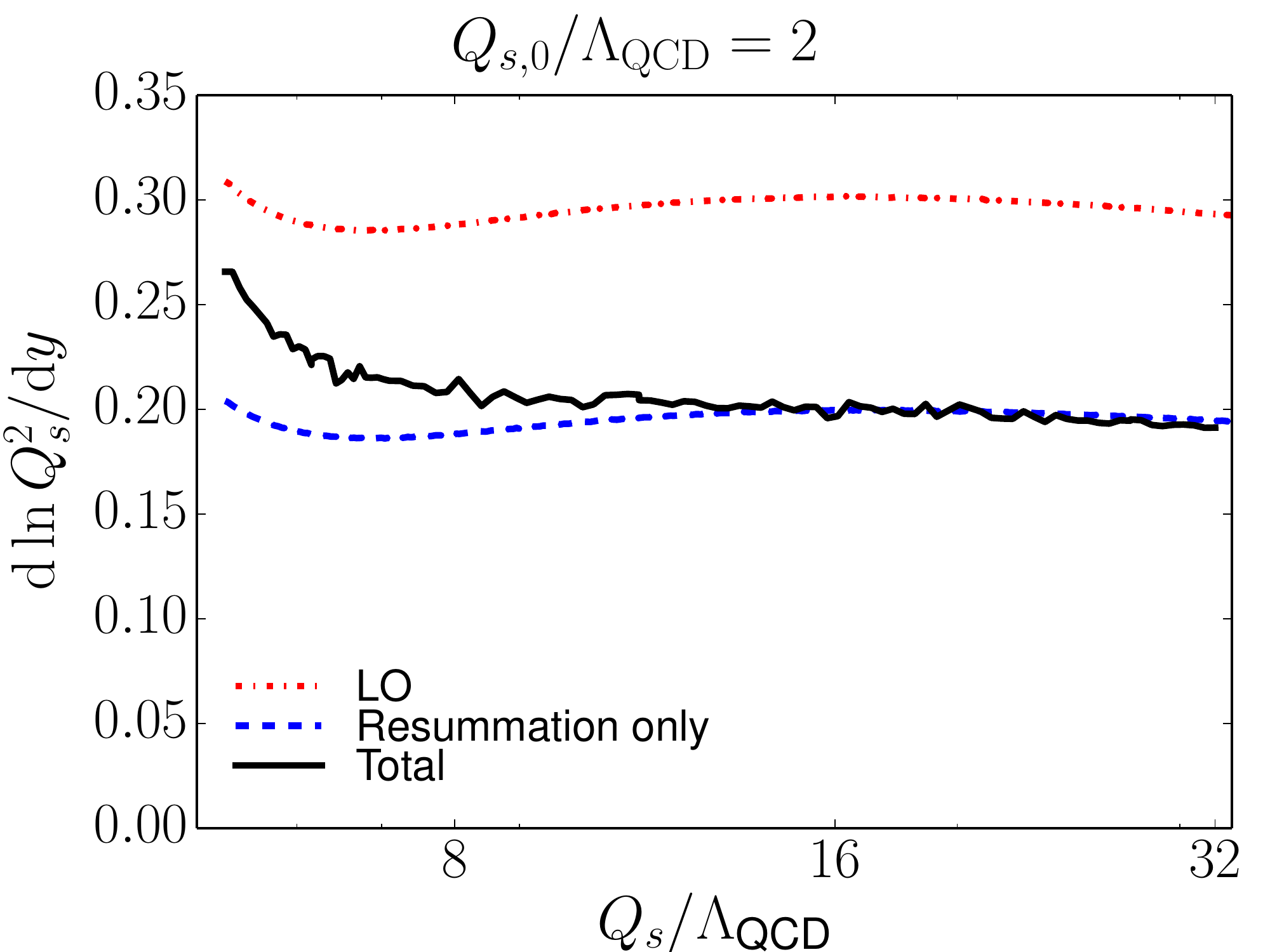}
\end{center}
\caption{
Evolution speed of the saturation scale obtained by solving the BK equation at leading order (with running coupling), including the resummation contributions and with full kernels with fixed order $\as^2$ terms. 
}\label{fig:lambda}
\end{figure}

\begin{figure*}[ptb]
	\subfloat[$\qso/\lqcd = 2$]{
		\includegraphics[width=0.45\textwidth]{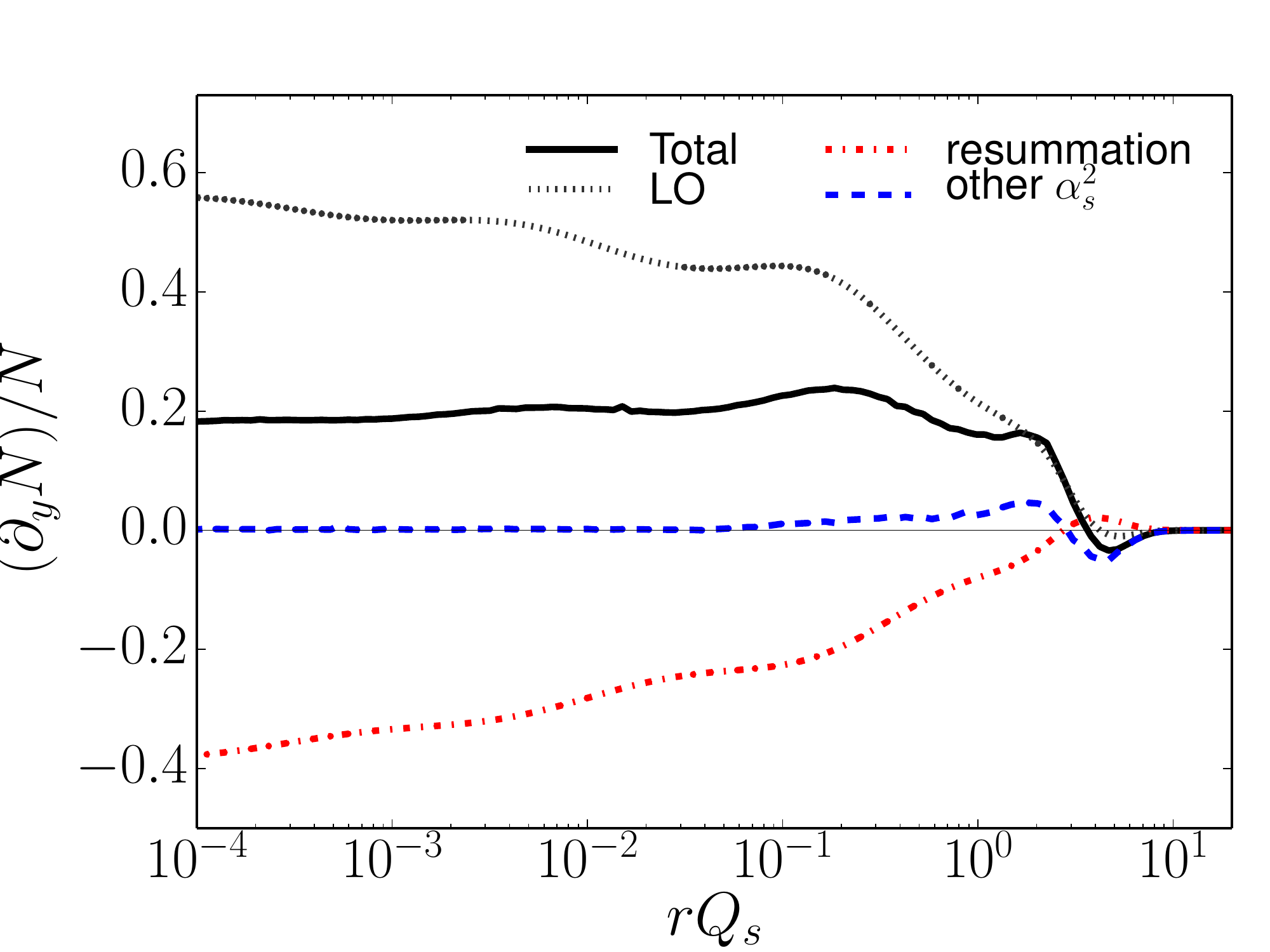}
		}
	\subfloat[$\qso/\lqcd=10$]{
	\includegraphics[width=0.45\textwidth]{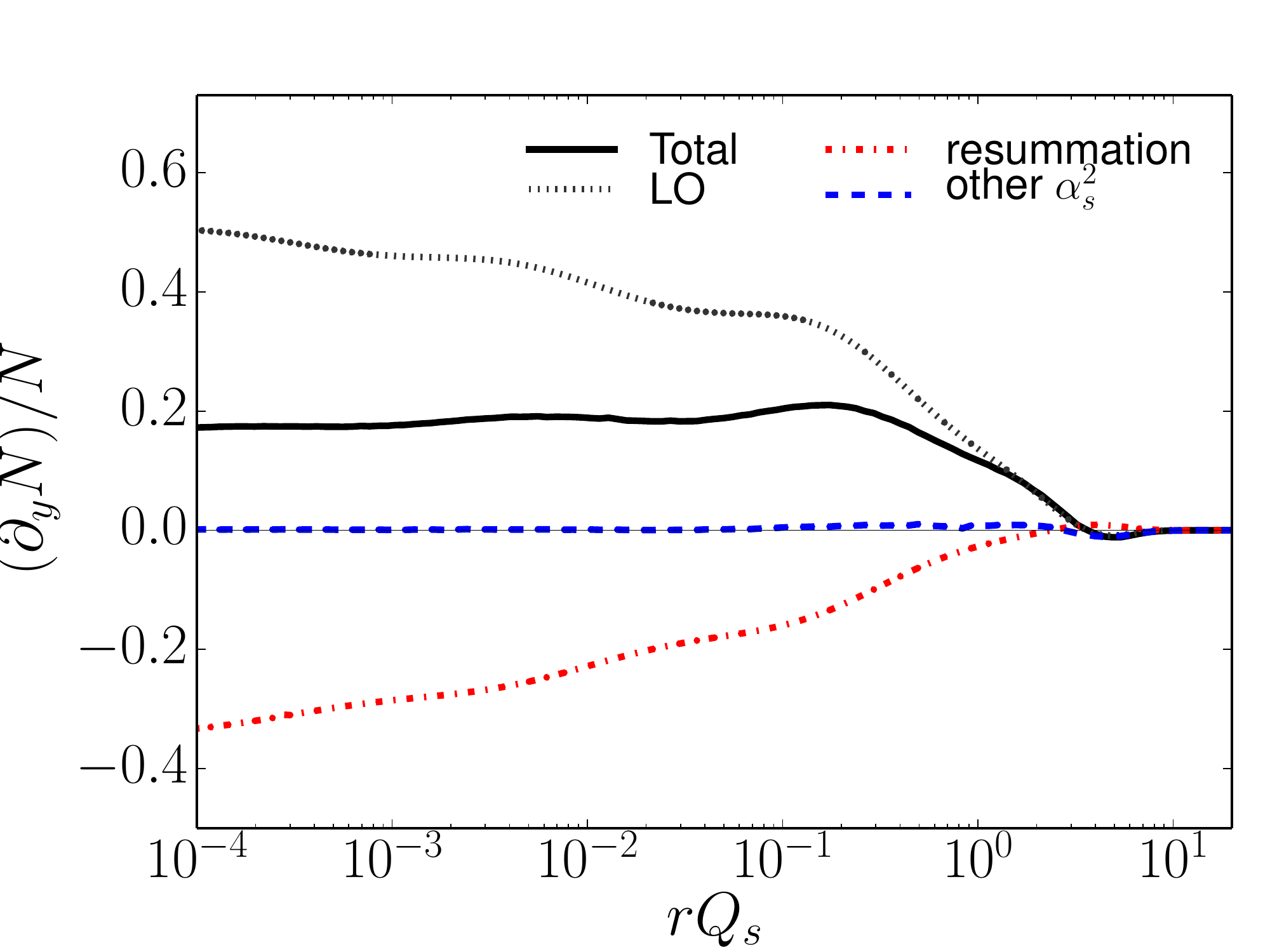}
		}
	\caption{Evolution speed of the dipole amplitude at the initial condition $y=0$ as a function of dipole size. The contributions from the leading order BK equation, resummation and the fixed order $\as^2$ terms are shown separately.}
	\label{fig:dndy_terms}
\end{figure*}

\begin{figure*}[ptb]
	\subfloat[$\qso/\lqcd = 2$]{
		\includegraphics[width=0.45\textwidth]{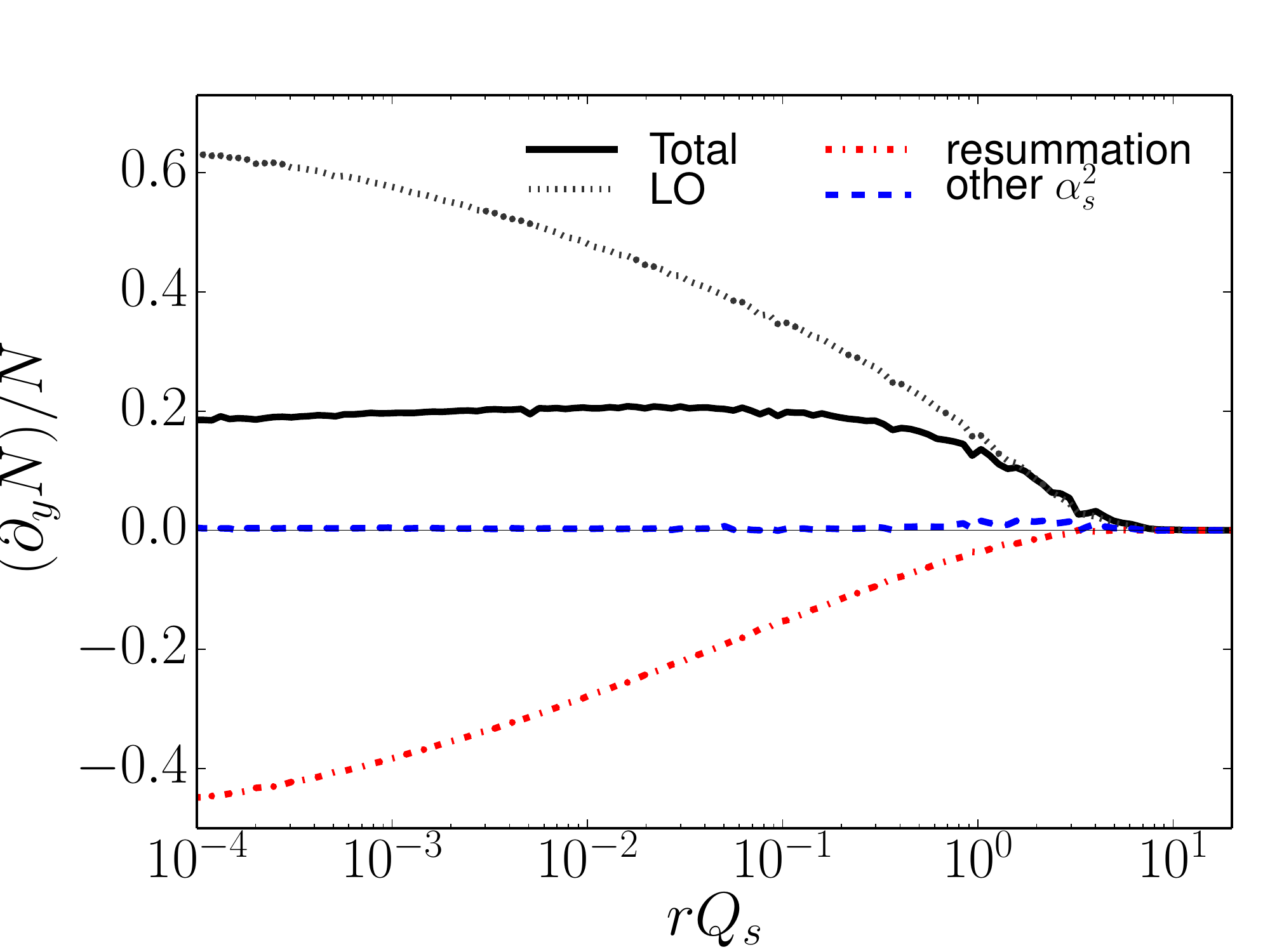}
		}
	\subfloat[$\qso/\lqcd=10$]{
	\includegraphics[width=0.45\textwidth]{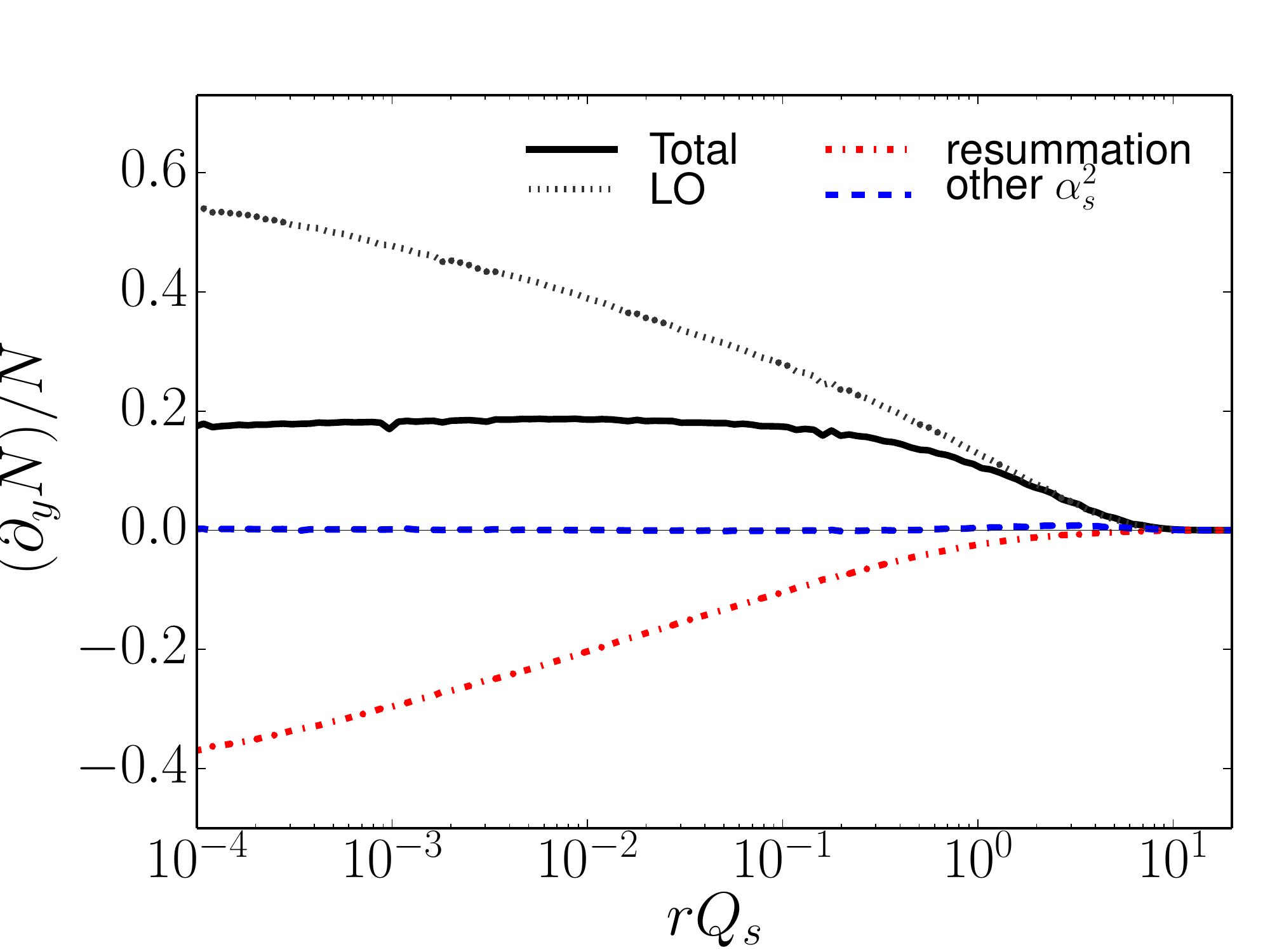}
		}
	\caption{Evolution speed of the dipole amplitude after the evolution at $y=10$ as a function of dipole size. The contributions from the leading order BK equation, resummation  and the fixed order $\as^2$ terms are shown separately.}
	\label{fig:dndy_terms_y_10}
\end{figure*}
The evolution of the saturation scale is studied in more detail in \fig\ref{fig:lambda} where we show its evolution speed $\der \ln \qs^2/\der y$. The saturation scale $\qs$ is defined here by
\begin{equation}
	N(r^2=2/\qs^2) = 1 - e^{-1/2},
\end{equation}
and it should be seen as the  scale at which non-linear phenomena become important.
The resummed NLO BK equation (Eq.~\eqref{eq:nlobk} with $K_1^\text{BC}$ replace by Eq.~\eqref{eq:full_k1}, labeled as \emph{Total}) is found to evolve roughly $30\%$ slower than the leading order running coupling BK equation at very large saturation scales with the running coupling prescription used here. 
The fixed order $\as^2$ terms are important close to the initial condition, increasing the evolution speed significantly. This can be seen by comparing the full resummed NLO BK result to the result obtained by solving the leading order BK equation improved as in Ref.~\cite{Iancu:2015joa} by including the resummation of single and double logarithms without the other NLO terms (\emph{Resummation only} in Fig.~\ref{fig:lambda}).
Later in the rapidity evolution (at large saturation scales) these pure NLO terms have a negligible effect. Note that we have here chosen an initial 
saturation scale $\qs \sim 1 \gev$, which can be expected to be in the phenomenologically relevant regime.

\begin{figure*}[ptb]
	\subfloat[$y=0$]{
		\includegraphics[width=0.45\textwidth]{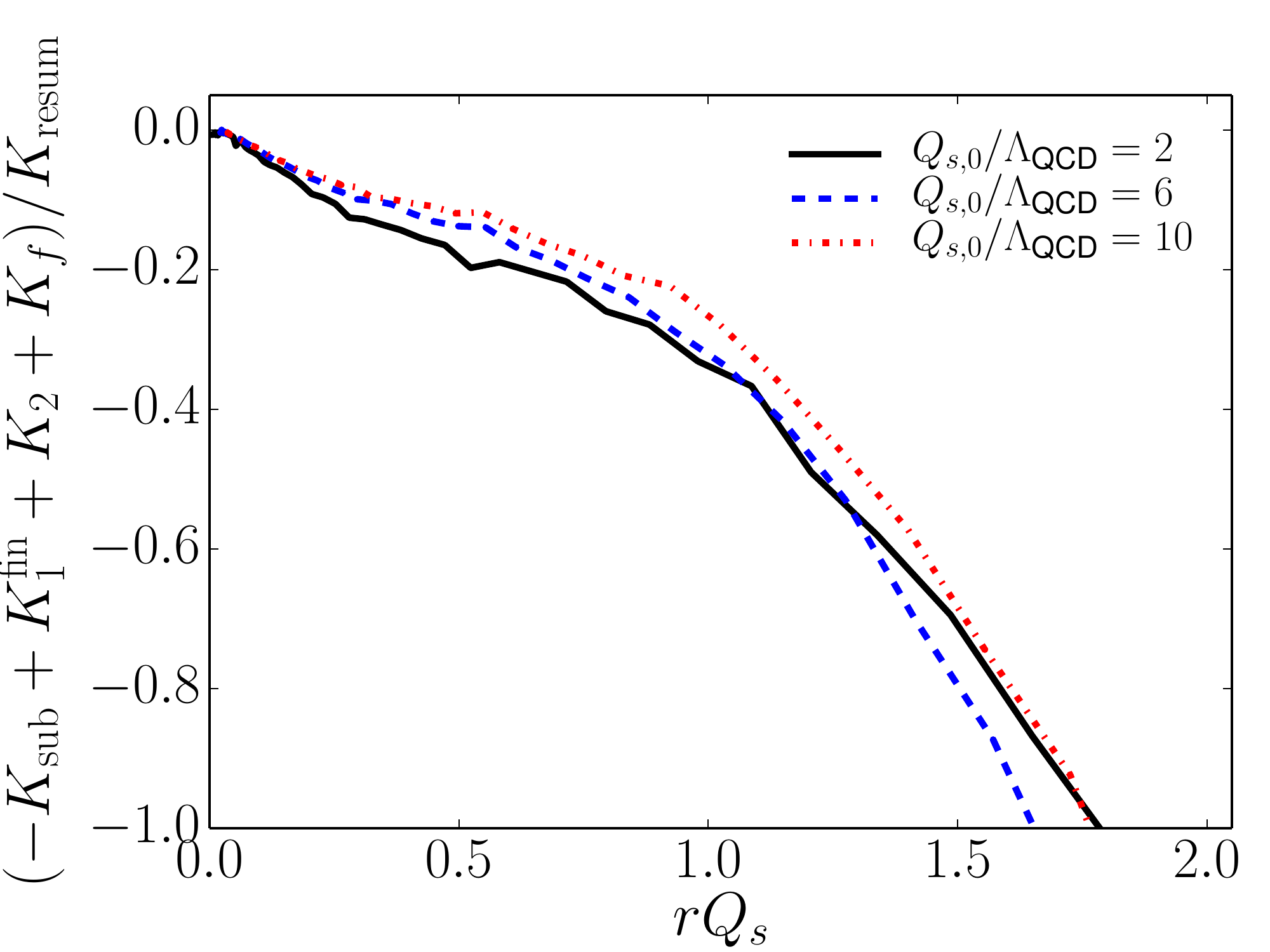}
		}
	\subfloat[$y=10$]{
	\includegraphics[width=0.45\textwidth]{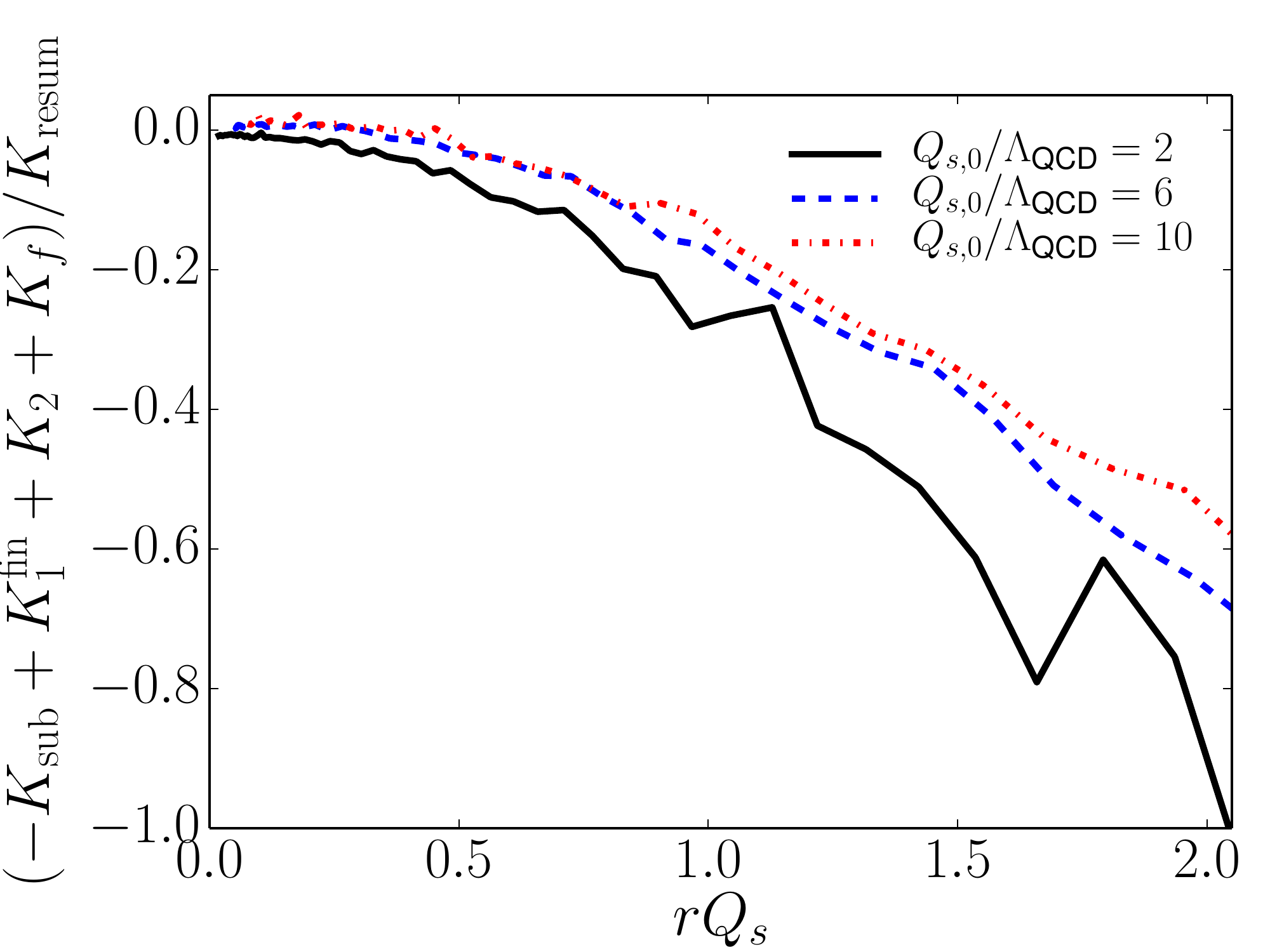}
		}
	\caption{Contribution to the evolution speed of the dipole amplitude from the $\as^2$ terms normalized by the corresponding contribution from the resummation. }
	\label{fig:dndy_as_resum_ratio}
\end{figure*}

The evolution speed of the dipole amplitude as a function of dipole size is analyzed in more detail in Fig.~\ref{fig:dndy_terms}, where the contributions to $\partial_y N(r)/N(r)$ from the different terms are shown. The resummation contribution is defined as
\begin{multline}
\label{eq:kresum}
K_\text{resum}=\frac{\as(r) \nc}{2\pi^2} \left(K_\text{DLA} K_\text{STL}-1\right) \\
\times \left[\frac{r^2}{X^2Y^2} + \frac{1}{X^2} \left(\frac{\as(X)}{\as(Y)}-1\right) + \frac{1}{Y^2} \left(\frac{\as(Y)}{\as(X)}-1\right) \right]  ,
\end{multline}
which is convoluted with the dipole part $D_1$. This corresponds to the contribution of the resummed NLO equation of~\cite{Iancu:2015joa} on top of the usual running coupling LO equation. The fixed order $\as^2$ contribution consists of the additional contribution of the kernels $K_\text{sub}$, $K_1^\text{fin}$, $K_2$ and $K_f$. We find that the fixed order NLO terms give a very small positive contribution to the evolution speed at small dipoles, and the resummed  large logarithms significantly slow down the evolution speed.
Note that while $K_2$ and $K_\text{sub}$ separately have a large single logarithmic contribution at small parent dipoles, this cancels in the total fixed order $\as^2$ term (``other $\as^2$'' in \fig\ref{fig:dndy_terms}).
 At larger dipoles $r\sim 1/\qs$ the resummation and the other NLO contributions are numerically equally important and mostly cancel each other, and the total evolution speed is close to the evolution of speed of the leading order BK equation in this regime.

When the calculation is done at larger saturation scales by increasing the value of $\qso$, the relative importance of fixed order $\as^2$ terms compared to the resummation around $r \sim 1/\qs$ is decreased. The same effect is observed when the contributions are studied after $10$ units of rapidity evolution in \fig\ref{fig:dndy_terms_y_10}. This corressponds to saturation scales $\qs/\lqcd \approx 19$ and $\qs/\lqcd \approx 66$ for $\qso/\lqcd=2$ and $\qso/\lqcd=10$, respectively.  The oscillations visible at $y=0$, that originate from the resummation of the initial condition, are washed out in the evolution.  It can also be seen that the MV model initial condition is closer to the asymptotic solution of the resummed NLO BK equation than it is for the leading order equation, as $\partial_y N/N$ is roughly constant in a much larger range of parent dipole sizes.
 
Let us  then demonstrate the importance of the fixed order $\as^2$ contributions at $r\sim 1/\qs$ relative to the resummation effects in more detail.  In \fig\ref{fig:dndy_as_resum_ratio} we show the contribution to the rapidity derivative of the dipole amplitude, $\partial_y N(r)$, originating from the $\as^2$ terms, normalized by the contribution of the resummation terms. That is, we show the ratio
\begin{equation}
\frac{-K_\text{sub} + K_1^\text{fin} + K_2 + K_f}{K_\text{resum}},
\end{equation}
where the resummation contribution $K_\text{resum}$ is defined in \eq\eqref{eq:kresum}, and all the kernels are convoluted with the corresponding dipole parts. 
As can be seen from \fig\ref{fig:dndy_as_resum_ratio} the resummation of single and double logarithms captures most of the higher-order corrections only at small dipoles. The fixed order $\as^2$ corrections become comparable to the resummation terms around $r\sim 1/\qs$, and their relative importance decreases in the evolution, as can be seen by comparing the calculations done at the initial condition and after $10$ units of rapidity evolution.

As shown previously in \re\cite{Lappi:2015fma} the NLO BK equation without resummation is very sensitive to the behaviour of the dipole amplitude at small dipoles, and with sufficiently steep small-$r$ slope the evolution turns unstable. To study this, we have solved the evolution equation with a set of initial conditions
\begin{equation}
	\label{eq:mvgamma}
	N(r)=1 - \exp\left[ -(r^2\qso^2)^\gamma \ln \left(\frac{1}{r\qso}+e\right)\right]
\end{equation}
varying the parameter $\gamma$ that controls the small-$r$ behavior. This parametrization is close to the MV$^\gamma$ model successfully fit to HERA deep inelastic scattering data with $\gamma \sim 1.1$ in \re\cite{Albacete:2010sy}. Note that the value $\gamma\sim 1.1$ is a result of a LO fit, and the phenomenologically relevant parameters for the NLO BK evolution are not necessarily the same.  In \re\cite{Lappi:2015fma} it was shown that the NLO BK equation becomes unstable at $\gamma \gtrsim 0.8\dots 1$.

\begin{figure}[tb]
\begin{center}
\includegraphics[width=0.49\textwidth]{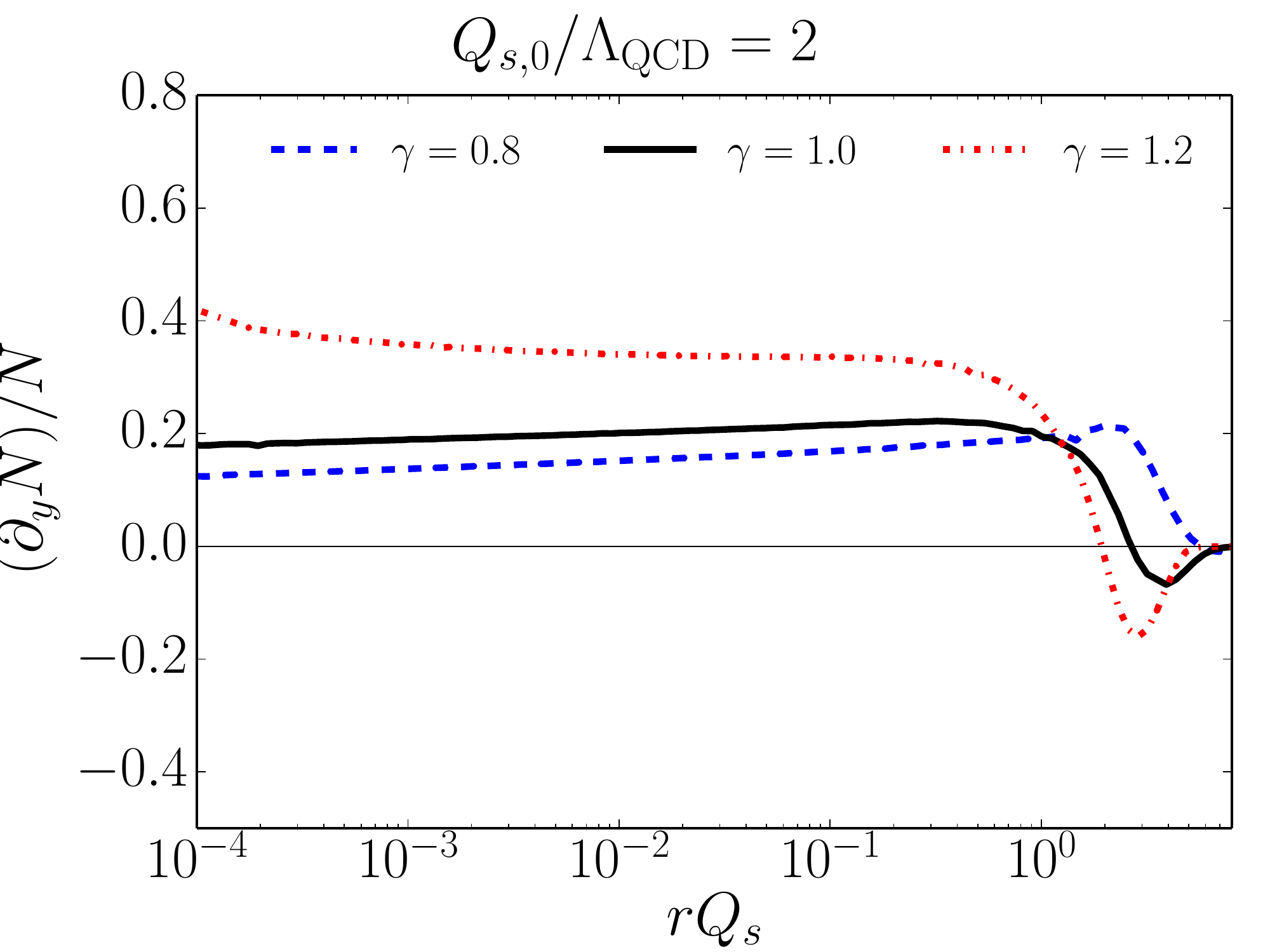}
\end{center}
\caption{
Evolution speed at $y=0$ using different values for the anomalous dimension at the initial condition MV$^\gamma$ parametrization, see \eq\eqref{eq:mvgamma}. 
}\label{fig:dndy_stability}
\end{figure}

The stability of the resummed NLO BK equation is studied by solving the equation using \eq\eqref{eq:mvgamma} as an initial condition with anomalous dimensions $\gamma=0.8,1.0$ and $\gamma=1.2$. As we are interested in the stability of the evolution equation only, the initial condition is not resummed.
 The obtained evolution speeds for the dipole amplitude $N(r)$ at the initial condition are shown in \fig\ref{fig:dndy_stability}. We find that with the resummed evolution equation a positive evolution speed at small dipoles is obtained with all values for the anomalous dimension $\gamma$, in contrast to the NLO BK equation without resummation of large logarithms.  Note that we use the same $C_\text{sub}=0.65$ when solving the NLO BK equation with an anomalous dimension in the initial condition even tough it is not exactly an optimal value for $\gamma\neq 1$.

\begin{figure}[tb]
\begin{center}
\includegraphics[width=0.49\textwidth]{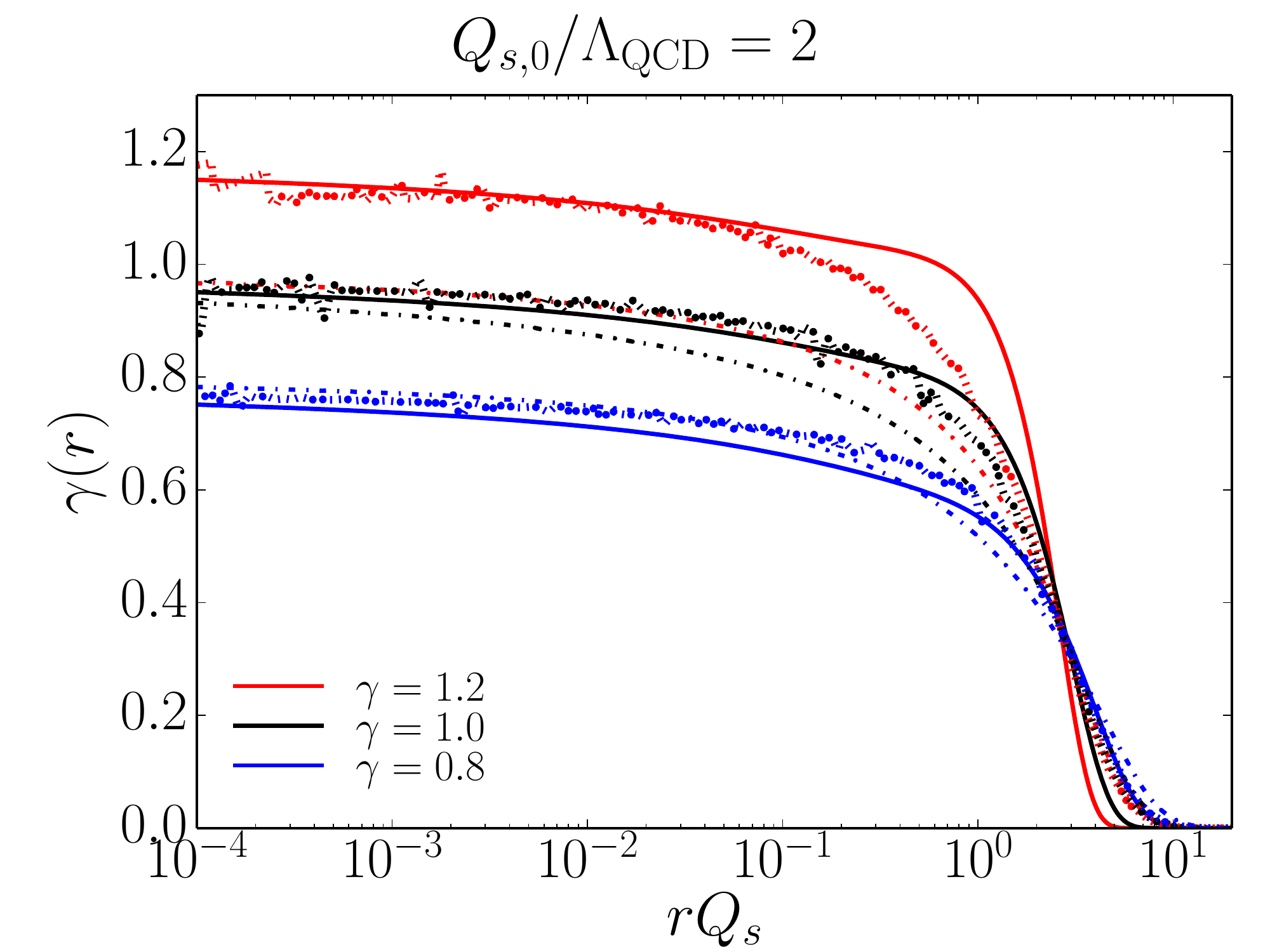}
\end{center}
\caption{
Anomalous dimension $\gamma(r)=\der \ln N(r)/ \der \ln r^2$ as a function of dipole size at the initial condition (solid lines) and after $5$ units of rapidity evolution (dotted lines). The initial conditions are the same as in Fig.~\ref{fig:dndy_stability}. For comparison, the leading order result at $y=5$ is shown as a dashed-dotted line.}
\label{fig:gamma_evol}
\end{figure}

To study how the shape of the dipole amplitude changes during the evolution we also calculate the anomalous dimension $\gamma(r)$ as a function of the parent dipole size. It is defined as
\begin{equation}
	\gamma(r) = \frac{ \der \ln N(r)}{\der \ln r^2}.
\end{equation}
The obtained anomalous dimension at the initial condition and after $5$ units of rapidity evolution are shown in Fig.~\ref{fig:gamma_evol}. For comparison the corresponding anomalous dimension obtained by solving the leading order BK equation with running coupling is shown. We find that the resummed NLO BK equation preserves the anomalous dimension of the initial condition, which suggest that the MV$^\gamma$ model parametrization is close to the asymptotic solution of the equation. On the other hand with leading order BK equation a significant rapidity evolution of $\gamma(r)$ is seen, especially with large anomalous dimension in the initial condition.

\section{Conclusions}

We have included the fixed order $\as^2$ corrections to the resummed Balitsky-Kovchegov evolution equation.
The main results of this work are presented in Figs.~\ref{fig:lambda} and~\ref{fig:dndy_as_resum_ratio}, where we show that at large saturation scales and at small dipoles the most important next-to-leading order corrections can be included in the BK equation by resumming large transverse logarithms. We have numerically found an optimal value for the constant inside the resummed logarithm that minimizes the effect of the other NLO terms. The fixed order $\as^2$ terms are numerically important close to the phenomenologically relevant initial conditions for large dipoles, $r\sim 1/\qs$, and significantly increase the evolution speed of the saturation scale. These terms become negligible at larger saturation scales (later in the evolution) and at small parent dipoles.

The resummed evolution equation is also shown to be stable and to generate physically meaningful evolution for the dipole amplitude even if an anomalous dimension $\gamma>1$ is used in the initial condition. This was not the case with the original NLO BK equation without resummation, as it was previously shown in \re\cite{Lappi:2015fma} to cause the dipole amplitude to turn negative with physically relevant initial conditions.

A logical next step towards the NLO CGC phenomenology would be to combine the resummed NLO BK evolution with the NLO photon impact factor~\cite{Balitsky:2010ze,Beuf:2011xd} and calculate the structure functions. In particular, the NLO CGC picture should be tested against the precise HERA deep inealstic scattering data~\cite{Aaron:2009aa,Abramowicz:2015mha}.

\section*{Acknowledgements} 
We thank D. Triantofyllopoulos and R. Paatelainen for discussions. This work has been supported by the Academy of Finland, projects 
267321 and 273464,
and by computing resources from
CSC -- IT Center for Science in Espoo, Finland.
H. M. is supported under DOE Contract No. DE-SC0012704.

\bibliography{../../../refs}
\bibliographystyle{JHEP-2modlong}

\end{document}